# The meaning of the *Sky of Salamanca*

## Astronomy versus Astrology

Carlos Tejero Prieto
Departamento de Matemáticas and Instituto de Física Fundamental y Matemáticas
Universidad de Salamanca
Plaza de la Merced, 1-4
37008 Salamanca

Abstract: After presenting the main characteristics of the *Sky of Salamanca*, we analyse whether what is represented in it is motivated by astrological or astronomical considerations. We will see that the astrological explanation based on the system of planetary houses described in Claudius Ptolemy's *Tetrabiblos* does not correspond to the planetary configuration we see in the salmantine vault. Finally, we will conclude that the astronomical interpretation of the Sky of Salamanca is justified by the representation of the Universe known in the 15th century according to the Almagest, by the placement of the stars in the constellations according to that work and by the dating of the painted planetary configuration, which circumstantial evidence places in August 1475.

### A question mark before our eyes

The *Sky of Salamanca* is a magnificent work of art in the Hispano-Flemish Gothic style that can be admired in the Patio de Escuelas Menores of the University of Salamanca. It formed part of the vault of the old library of the Salamanca *Studium*[1], which was located in the place now occupied by the Royal Chapel of St. Jerome in the Escuelas Mayores building. After a series of misfortunes, two thirds of the original vault were lost in the 18th century, and the part that remains today, after being removed from its original location by means of the *strappo* technique and later restored, was moved to the place it occupies today at the end of 1952.

We know from the writings of Hieronymus Münzer, Lucio Marineo Sículo and Diego Pérez de Mesa [6-10], we know that the 48 constellations described by Claudius Ptolemy in the *Almagest* [11, Books VII, VIII] were represented in the original vault, as well as the 7 planets known in the 15th century: Mercury, Venus, Mars, Jupiter and Saturn alongside with the Sun and the Moon, which were considered planets at the time.

At the time of the construction of the vault, Astrology comprised both Natural Astrology, which corresponds to what we now call Astronomy, and Judicial Astrology, which is the pseudo-science we now call simply Astrology and which deals with the supposed influence of the celestial bodies on human events. In what follows, we shall use the current terminology of Astronomy and Astrology in the sense just described.

The *Almagest* [11] was the standard treatise in astronomy for 1400 years, from its writing in the 2nd century until the end of the 16th century when it began to be slowly displaced by Nicolaus Copernicus' *De revolutionibus orbium coelestium* [2].

But Claudius Ptolemy also wrote the fundamental work on Astrology, called in Greek *Apotelesmatiká* [14]. Since this treatise was divided into four books, it is also known as the *Tetrabiblos* in its Greek root [13, 14] or as the *Quadripartitum* in its Latin translations.

---

[1] The library was built between 1474 and 1479 and was most probably decorated between 1483 and 1486 [8,9]. Its vault was 23 metres long, 8.70 metres wide and 4.35 metres high from its starting point.

Both the Almagest and the Tetrabiblos represented the state of astronomical and astrological knowledge, respectively, at the time of the construction of the old university library. As Noehles-Doerk [5] and Martínez Frías [9] have pointed out, the decision to enclose the library with a vault is an unmistakable sign that it had already been decided to decorate it before beginning its construction, and therefore its iconographic programme had been planned before the works were undertaken. The decoration that finally took shape on the ceiling of the library is testimony to the importance that the chair of Astrology at the University of Salamanca had soon acquired, since Nicolás Polonio was its first incumbent between 1460 and 1464.

It is therefore natural to ask whether the pictorial work done in the ancient library corresponded to astrological or astronomical aspects. The answer, which may be interdisciplinary, must in any case explain what we see represented in the *Sky of Salamanca*. In Fig. 1 we see the four boreal constellations (Boötes, Hercules, Ophiucus and Serpens), five zodiacal constellations (Leo, Virgo, Libra, Scorpio and Sagittarius), seven austral constellations (Hydra, Crater, Corvus, Centaurus, Lupus, Ara and Corona Australis) and two planets (the Sun and Mercury).

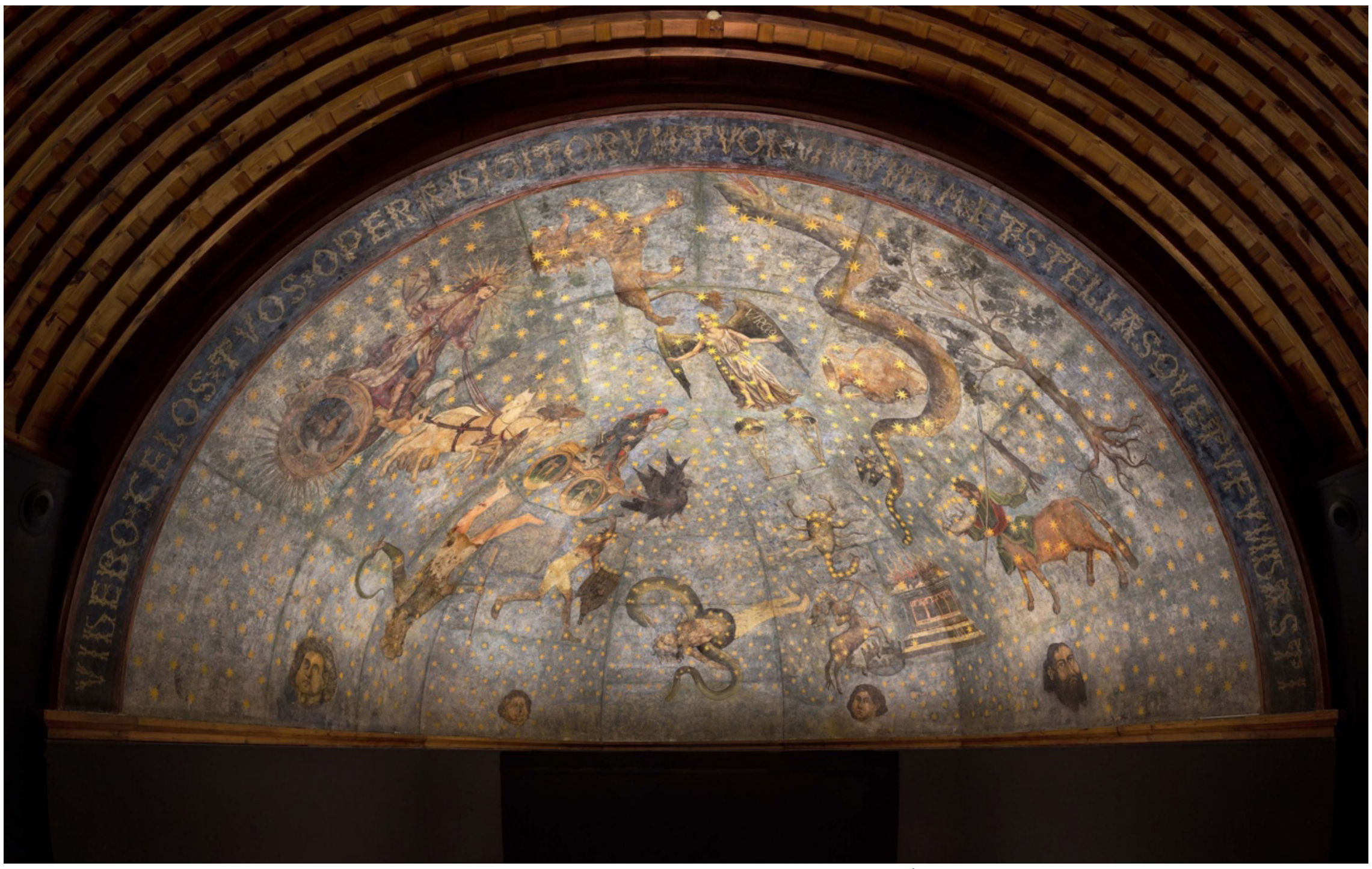

*Fig. 1 El Cielo de Salamanca. Universidad de Salamanca / Santiago Santos.*

In 1959, Ernst Zinner[2] observed the Salamanca sky in situ as part of a research trip to Spain and Portugal. The results of his visit were reported in [17], where he concludes that what we can observe in the preserved part of the Salamanca vault is a type of representation of Italian influence, which was fashionable at the time of its decoration, in which the planets appear personified as rulers on triumphal chariots. From Zinner's study we conclude that the Sun is in the constellation Leo and Mercury in the constellation Virgo, but without being able to further specify their location as the main celestial circles are not depicted. Later researchers, such as Noehles-Doerk [5], the astronomer Duerbeck [5] and Esteban Lorente [3] continued this line of research, see also [15].

[2] E. Zinner was one of the 20th century's greatest experts on the history of astronomy focusing especially on the Renaissance period [16]. He published more than 350 research papers [18] of which 9000 pages are devoted to his historical studies. He was director of the Remeis Astronomical Observatory in Bamberg, Germany, for almost 30 years, member of the International Astronomical Union (Commission 41, History of Astronomy), of the German Academy of Natural Sciences Leopoldina in Halle, of the Académie Internationale d'Histoire des Sciences, of the German Society for the History of Medicine, Natural Sciences and Technology and of the Society for Natural Research in Bamberg, Leibniz Medal of the Prussian Academy of Sciences in Berlin, etc. For more information on his scientific profile see [16].

## The impossible astrological connection

Could it be that the iconographic programme represented in the ancient salmantine library had a motivation of astrological origin? To try to answer this question we will make use of Ptolemy's *Tetrabiblos* which, as we have already said, was the reference work in this field in the 15th century.

Since all the then known planets were represented in the vault and they move along the band of zodiacal constellations, let us see what the *Tetrabiblos* has to say about them.

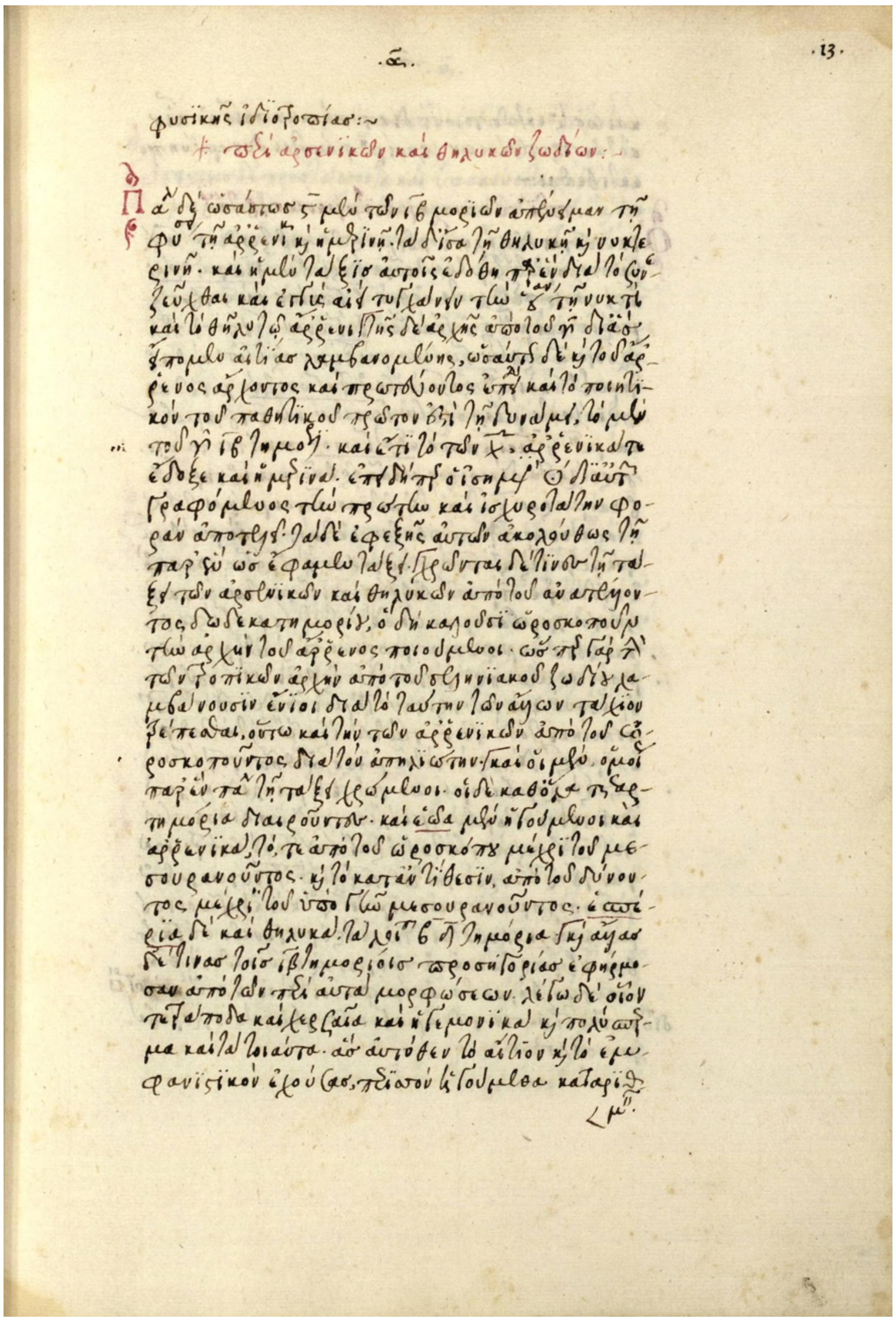

*Fig. 2 Tetrabiblos, Book I, 12 [14, f. 13r]. Universidad Complutense.*

In Book I, we have [13, pág. 69]:

*12. Of Masculine and Feminine Signs.*

*Again, in the same way they assigned six of the signs to the masculine and diurnal nature and an equal number to the feminine and nocturnal. An alternating order was assigned to them because day is always yoked to night and close to it, and female to male. Now as Aries is taken as the starting-point for the reasons we have mentioned, and as the male likewise rules and holds first place, since also the active is always superior to the passive in power, the signs of Aries and Libra were thought to be masculine and diurnal, an additional reason being that the equinoctial circle which is drawn through them completes the primary and most powerful movement of the whole universe. The signs in succession after them correspond, as we said, in alternating order.*

Thus, following this prescription we have the following graphical representation of the distribution of diurnal and nocturnal zodiacal signs:

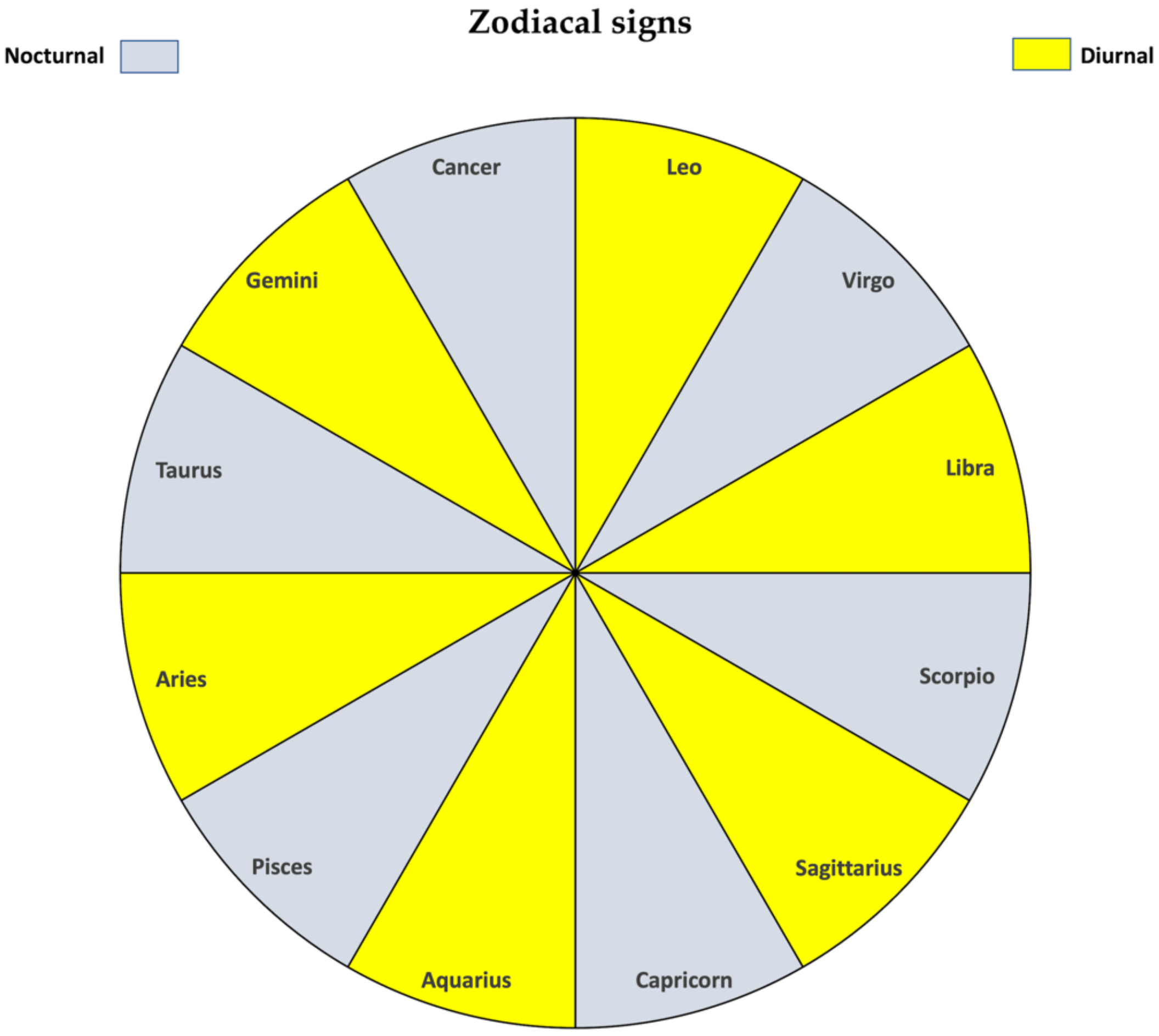

*Fig. 3 Diurnal and nocturnal zodiacal signs according to the Tetrabiblos.*

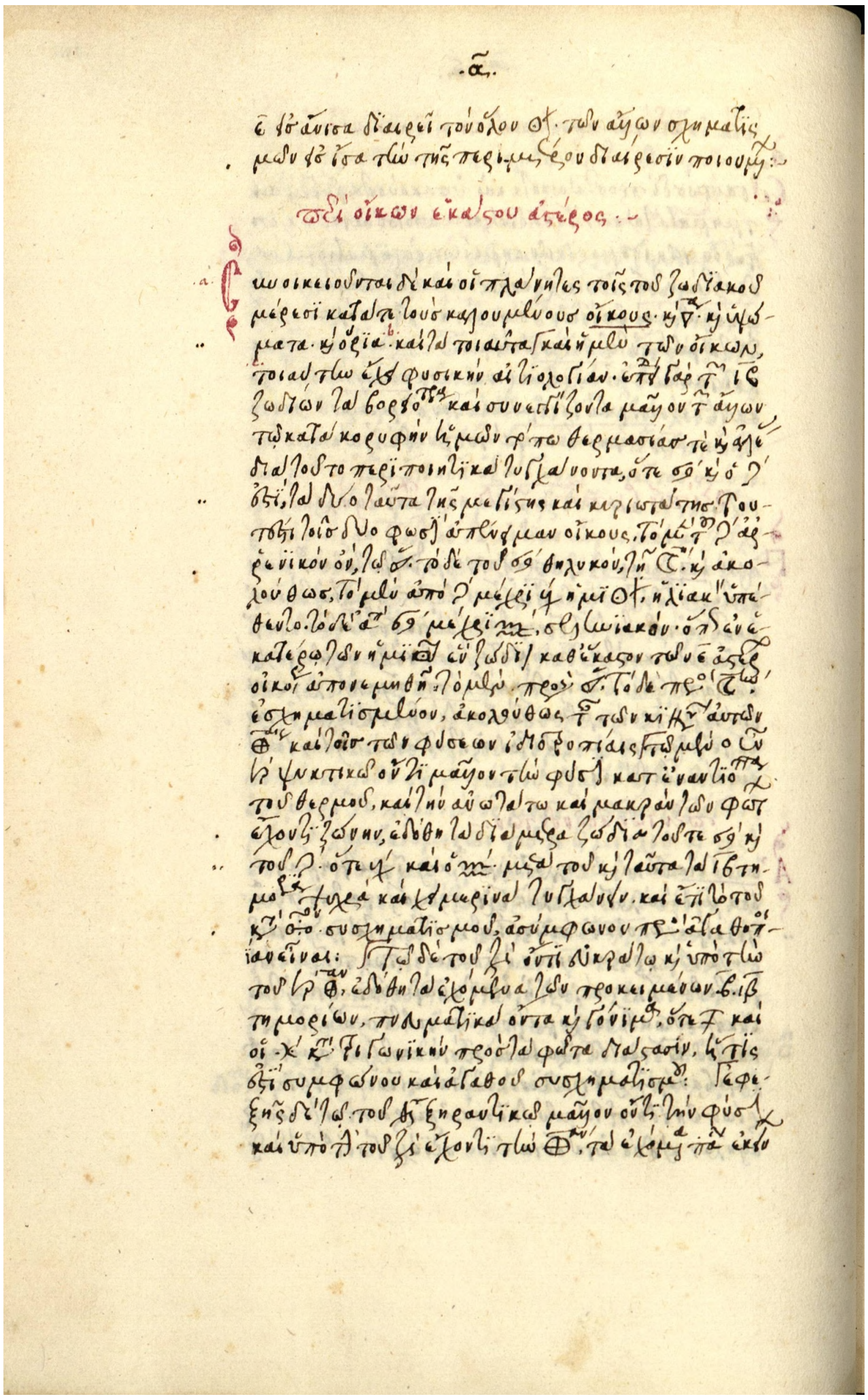

*Fig. 4 Tetrabiblos, Book I, 17 [14, f. 14v]. Universidad Complutense.*

Subsequently, in the same Book I of the *Tetrabiblos* [13, pág. 79] one reads that:

*17. Of the Houses of the Several Planets.*

*The planets also have familiarity with the parts of the zodiac, through what are called their houses, triangles, exaltations, terms, and the like. The system of houses is of the following nature. Since of the twelve signs the most northern, which are closer than the others to our zenith and therefore most productive of heat and of warmth are Cancer and Leo, they assigned these to the greatest and most powerful heavenly bodies, that is, to the luminaries, as houses, Leo, which is masculine, to the sun and Cancer, feminine, to the moon. In keeping with this they assumed the semicircle from Leo to Capricorn to be solar and that from Aquarius to Cancer to be lunar, so that in each of the semicircles one sign might be assigned to each of the five planets as its own, One bearing aspect to the sun and the other to the moon, consistently with the spheres of their motion and the peculiarities of their natures. For to Saturn, in whose nature cold prevails, as opposed to heat, and which occupies the orbit highest and farthest from the luminaries, were assigned the signs opposite Cancer and Leo, namely Capricorn and Aquarius, with the additional reason that these signs are cold and wintry, and further that their diametrical aspect is not consistent with beneficence. To Jupiter, which is moderate and below*

*Saturn's sphere, were assigned the two signs next to the foregoing, windy and fecund, Sagittarius and Pisces, in triangular aspect to the luminaries, which is a harmonious and beneficent configuration. Next, to Mars, which is dry in nature and occupies a sphere under that of Jupiter, there were assigned again the two signs, contiguous to the former, Scorpio and Aries, having a similar nature, and, agreeably to Mars' destructive and inharmonious quality, in quartile aspect to the luminaries. To Venus, which is temperate and beneath Mars, were given the next two signs, which are extremely fertile, Libra and Taurus. These preserve the harmony of the sextile aspect; another reason is that this planet at most is never more than two signs removed from the sun in either direction. Finally, there were given to Mercury, which never is farther removed from the sun than one sign in either direction and is beneath the others and closer in a way to both of the luminaries, the remaining signs, Gemini and Virgo, which are next to the houses of the luminaries.*

Therefore, completing Fig. 2 with the information we have just obtained, the planetary houses are ordered as follows:

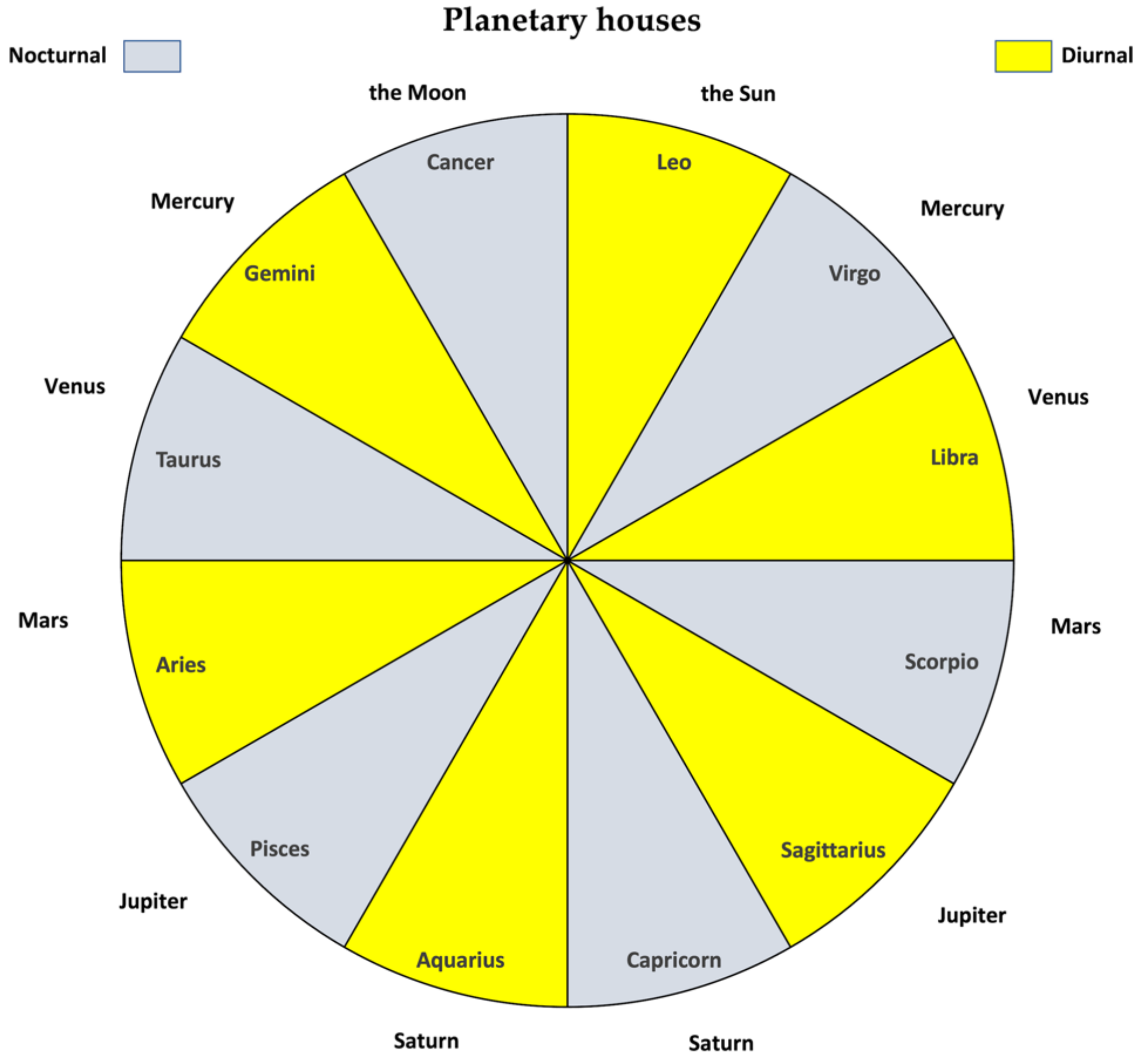

*Fig. 5 Diurnal and nocturnal planetary houses according to the Tetrabiblos.*

The astrological hypothesis of the planetary houses to explain what we see represented in the *Sky of Salamanca* would consist in affirming that the planets with only one house would be painted next to it, while those with two houses would be either all adjacent to their diurnal houses or all adjacent to their nocturnal houses. However, since in the preserved part of the vault Mercury is in its nocturnal house, the astrological hypothesis would imply that all planets with two houses should necessarily be depicted next to their nocturnal houses.

Now, taking into account the astrological information we have just collected in Figs. 4 and 5, we can conclude that the representation we see in the *Sky of Salamanca* cannot be explained by the system of planetary houses described in the *Tetrabiblos*. Let us see why this is so.

For planets with only one house there are no problems. In fact, the Sun is seen next to its only house, which is Leo, and the Moon could have been next to its only house, Cancer, and that is why we would not see it represented today in the *Sky of Salamanca*.

Could it be that the rest of the planets that have two houses have also been represented in their nocturnal houses and that is why they do not appear in the *Sky of Salamanca*? The definitive answer is no, since it is enough to look at Fig. 5 to realise that, if this were the case, then we should see in Fig. 1 the triumphal chariot of Mars next to Scorpio, which is its nocturnal house, and this is not the case.

Thus, we can state categorically that the planetary configuration that we see represented in the *Sky of Salamanca* is not a consequence of the arrangement of the planets that would be obtained from the astrological system of planetary houses described in the *Tetrabiblos*.

**The astronomical connection**

Diego Pérez de Mesa in his *Libro de las grandezas y cosas notables de España*[3] published in 1590 [10], describes that in the vault of the old university library there were:

> *"painted, and carved in gold the forty-eight images of the eighth sphere …"*

In other words, what was depicted conformed to the stellar catalogue described by Ptolemy in the Almagest, which consisted of 1028 stars grouped in forty-eight constellations forming the eighth sphere of fixed stars.

Thanks to the description of Lucio Marineo Siculo in his work *De Hispaniae Laudibus* [7] we are informed that the seven planets known at the time were also represented, see [6,9]. Thus, in the vault, the Universe as it was then understood was presented as described in the *Almagest*, which was the astronomical reference work.

The adoption of the iconography of the planets as rulers on triumphal chariots introduced, due to its spatial extension, obvious limitations in the composition. Thus, the painter had to adopt a compromise between figurative fidelity and the chosen iconographic programme. From what we see depicted in the *Sky of Salamanca*, we can infer that in the ancient vault, a qualitative representation of the celestial sphere was painted in which the relative positions of the constellations were preserved, while the planets, due to their size, were placed adjacent to the zodiacal band in the hemisphere with the lowest local density of constellations, determined by the type of projection used.

---

[3] An early version of this book, entitled *Libro de las grandezas y cosas memorables de España*, was published in 1543 in Seville by Pedro de Medina.

As Professor Flórez Miguel [4] pointed out, a planetarium was painted on the ceiling of the old library, the first and fundamental aim of which was the practical teaching of astronomy. It is the first of its kind in the history of astronomy that has come down to us, in the remaining part which we now call the *Sky of Salamanca*.

In the star catalogue of the *Almagest* [11, Books VII, VIII], [12, ΒΙΒΛΊΟΝ ΈΒΔΟΜΟΝ, ΒΙΒΛΊΟΝ “ΟΚΔΌΟΝ] the stars forming each constellation are described in words, including their ecliptic longitude and latitude and magnitude, but without any graphic representation. Therefore, throughout history the constellations have been represented in many different ways.

The stars at the bottom of the vault seem to be evenly distributed in order to simulate the night sky, without necessarily representing real stars. However, it is a remarkable fact, and unpublished until now, according to the sources consulted, that the stars of the constellations mostly correspond to the literal description given in the *Almagest*, although they do not adhere to the ecliptic longitude and latitude. For example, in the case of the constellation Leo (Fig. 7), the 14 stars painted in the *Sky of Salamanca* correspond to the *Almagest* description (Figs. 6 and 8), both in their placement and magnitude.

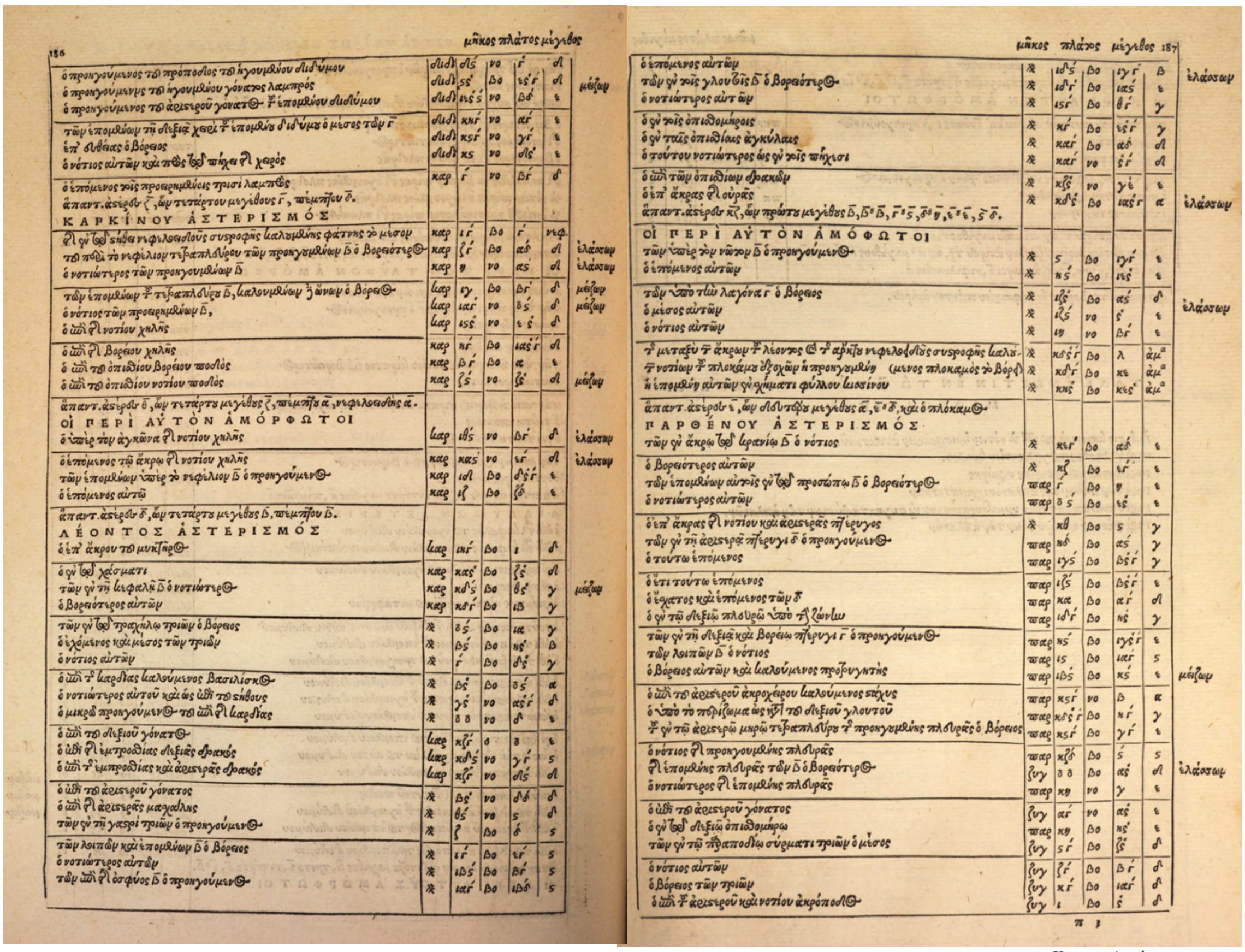

*Fig. 6 Description of the constellation Leo in Chapter VII of the Almagest [12, f. 186v, f. 186r]. Bayerische Staatsbibliothek.*

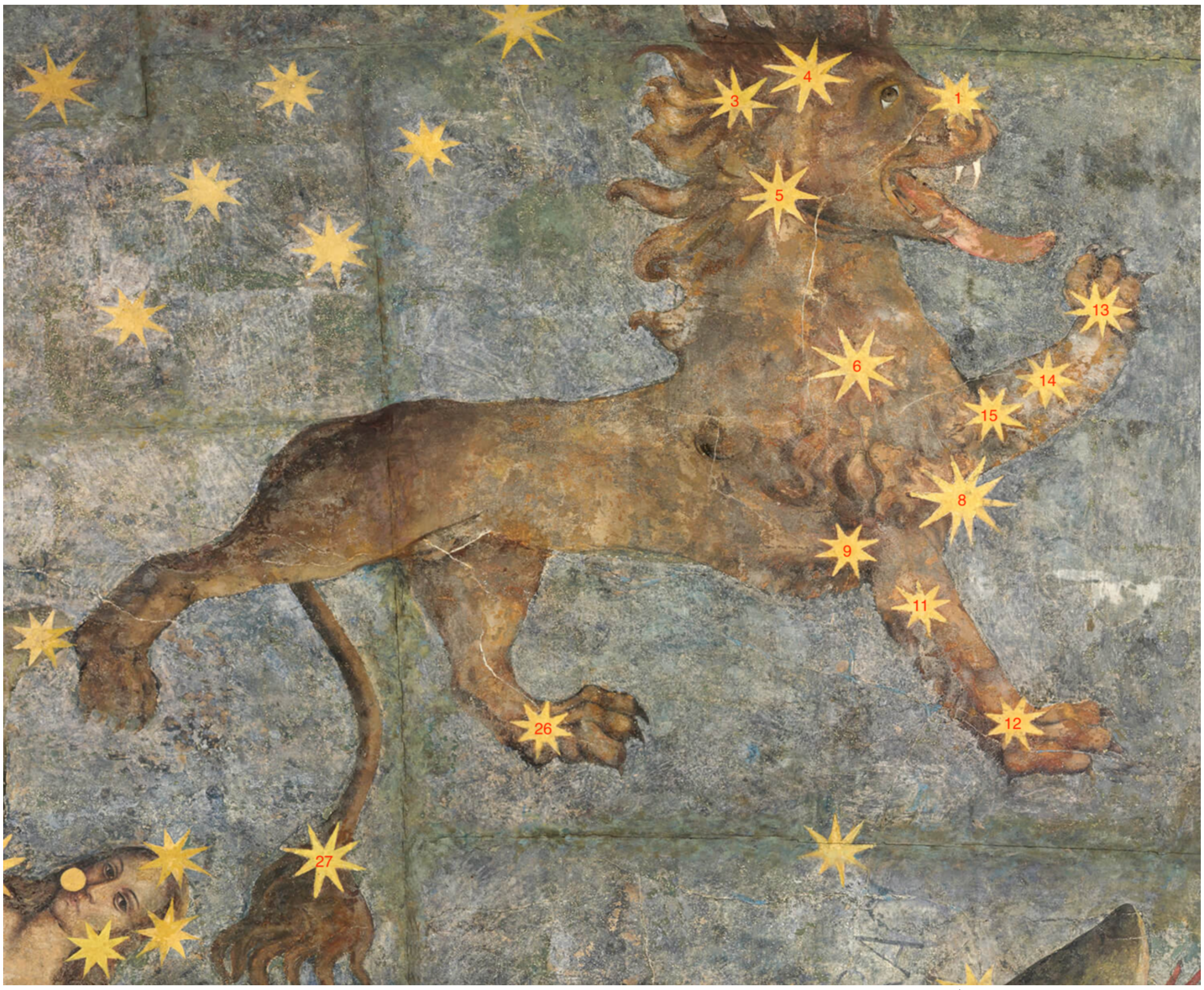

*Fig. 7 Numbering of the stars of the constellation Leo according to the Almagest. USAL / Carlos Tejero*

**Constellation of Leo**

1 The star on the tip of the nostrils
2 The star in the gaping jaws
3 The northernmost of the two stars in the head
4 The southernmost of these
5 The northernmost of the 3 stars in the neck
6 The one close to this, the middle one of the three
7 The southernmost of them
8 The star on the heart, called `Regulus'
9 The one south of this, about on the chest
10 The star a little in advance of the star on the heart [no. 8]
11 The star on the right knee
12 The star on the right front claw-clutch
13 The star on the left front claw-clutch
14 The star on the left [front] knee
15 The star on the left armpit
16 The most advanced of the three stars in the belly
17 The northernmost of the other, rearmost 2
18 The southernmost of these [two]
19 The more advanced of the two stars on the rump
20 The rearmost of them
21 The northernmost of the 2 stars in the buttocks
22 The southernmost of them
23 The star in the hind thighs
24 The star in the hind legs-bends
25 The one south of this, about in the lower legs
26 The star on the hind claw-clutches
27 The star on the end of the tail

**Stars around Leo outside the constellation:**

28 The more advanced of the 2 over the back
29 The rearmost of them
30 The northernmost of the 3 under the flank
31 The middle one of these
32 The southernmost of them
33 The northernmost part of the nebulous mass between the edges of Leo and Ursa [Major], called Coma [Berenices]
34 The most advanced of the southern outrunners of Coma
35 The rearmost of them, shaped like an ivy leaf

*Fig. 8 Description of the stars of the constellation Leo in the Almagest. [11, Book VII, págs. 367-368]*

The painter, very likely to facilitate the visibility of the image of the Lion, decided to depict only 14 of the 35 stars listed in the Almagest for the constellation Leo. The coincidence of the stars depicted in the *Sky of Salamanca* with their description in the *Almagest* clearly serves the purpose of the practical teaching of astronomy. However, this does not mean that what is represented in the vault is an exact transcription of the constellations as they are seen in the night sky, for as we have said before, the image that has been associated with the constellations has varied throughout history. The fact that it does not conform to the ecliptic references may be due to the possible use of the Spanish translation of the Almagest's star catalogue, contained in the *Libro de la ochava esfera* [1] belonging to the *Libros del saber de astrología* written under the patronage of King Alfonso X the Wise. In this Spanish version, neither the latitudes nor the ecliptic longitudes of the stars are included.

As already pointed out by Zinner in [17], in the *Sky of Salamanca* there is no doubt in the indication of a point in time by the position of the planets in the zodiac. Thus, it is key to determine when the planetary configuration depicted there can be observed, see Fig. 1. The years and days in which it can be seen are called the Years and Days of the Sky of Salamanca. In the previous work [15] I analysed the 1100 year period between 1200 AD. and 2300 AD., showing that there are only 23 such years, so they are extremely rare on a human scale.

Therefore, in principle, it would not be possible to assign a single date to the firmament represented in the old university vault. However, taking into account the extremely infrequent years of the *Sky of Salamanca*, it is an exceptional fact that in the short period of construction of the old library and the decoration of its vault - comprised in the period between 1474 and 1486 - there is one and only one of those years, namely 1475, in which the configuration could be observed from the 15th to the 28$^{th}$ of August. Thus, this circumstantial evidence seems to place the dating of the Salamanca work in that interval [15].

**Conclusion**

In view of the above, everything seems to indicate that the process of decorating the vault of the former university library began with the painter's own artistic criteria for the figurative programme of the constellations. Later, with the assistance of an astronomer, a suitable number of stars were selected and placed above the figures of the constellations in the places described in the Almagest. Even so, in several constellations there are discrepancies in the laterality and placement of some of their stars. These may be due in part to the modifications introduced by Juan de Yprés in 1506 when he removed the plaster relief of the constellation stars, eliminated others and proceeded to a general repainting of the entire work, which caused great damage [8]. It is plausible that the constellations located at the zenith of the vault, such as Leo, suffered fewer modifications due to the greater difficulty of accessing them because of their great height.

The decoration of the ancient university library at Salamanca was a pioneering work of scientific art, representing a planetarium in accordance with the knowledge of the time, whose main purpose was the teaching of astronomy. Moreover, it is a unique work in European university history of the 15th century, especially if we consider the unprecedented nature of a three-dimensional representation of the celestial sphere on this scale, with the difficulties inherent in calculating how to project the figures onto the vault, at a time when the development of plane cartography, and in particular celestial cartography, was still in its incipient stages.

For all these reasons, there is an unquestionable link between the *Sky of Salamanca* and hence the vault of the old university library with the astronomical model set out by Claudius Ptolemy in the *Almagest*. Moreover, an explanation cannot be provided by the astrological hypothesis based on the system of planetary houses described by the same author in the *Tetrabiblos*.